\documentclass[twoside,twocolumn,english]{revtex4}
\usepackage[T1]{fontenc}
\usepackage[latin1]{inputenc}
\usepackage{geometry}
\usepackage{graphicx}

\makeatletter

\usepackage{babel}
\makeatother
\begin{document}

\title{Crossover behavior in a mixed-mode fiber bundle model}

\author{Srutarshi Pradhan$^{a}$}

\email{pradhan.srutarshi@phys.ntnu.no}

\author{Bikas K. Chakrabarti$^{b}$}

\email{bikas@cmp.saha.ernet.in}

\author{Alex Hansen$^{a}$}

\email{alex.hansen@phys.ntnu.no}

\affiliation{$^{a}$Department of Physics, Norwegian University of Science and
Technology , Trondheim 7491, Norway}

\affiliation{$^{b}$Condensed Matter Physics Group, Saha Institute of Nuclear
Physics, 1/AF , Bidhan Nagar, Kolkata -700 064, India}

\begin{abstract}
\noindent We introduce a mixed-mode load sharing scheme in fiber bundle
model. This model reduces exactly to equal load sharing (ELS) and
local load sharing (LLS) models at the two extreme limits of a single
load sharing parameter. We identify two distinct regimes: a) Mean-field
regime where ELS mode dominates and b) short range regime dominated
by LLS mode. The crossover behavior is explored through the numerical
study of strength variation, the avalanche statistics, susceptibility
and relaxation time variations, the correlations among the broken
fibers and their cluster analysis. Analyzing the moments of the cluster
size distributions we locate the crossover point of these regimes.
We thus conclude that even in one dimension, fiber bundle model shows
crossover behavior from mean-field to short range interactions. 
\end{abstract}
\maketitle

\section{Introduction}

Fiber bundle model represents a simple, stochastic fracture-failure
process \cite{books} in materials subjected to external load. The
model consists of three basic ingredients: (a) a discrete set of $N$
elements located at sites of a lattice (b) a probability distribution
of the strength threshold of individual elements (fibers) (c) a load-transfer
rule which distributes the terminal load carried by the failed fibers
to the surviving fibers. The model study was initiated by Peirce \cite{Peirce}
in the context of testing the strength of cotton yarns. Since then
this model has been studied and modified by many groups {[}3-25{]}
using analytic as well as numerical methods. Fiber bundles are of
two classes with respect to the time dependence of fiber strength
threshold: `Static' bundles contain fibers whose threshold strengths
are independent of time and such bundles are subjected to quasi static
loading, i.e., the load is increased steadily up to the complete failure
of the bundles. The load or stress $\sigma$ (load per fiber) is an
independent variable here and the strength of the bundle is determined
by the maximum value of the applied load or stress ($\sigma_{c}$)
that can be supported by the bundle. On the other hand `dynamic' bundles
consist of fibers having time dependent strength and the fibers fail
due to fatigue \cite{BD-58,Phoenix00,Roux-00,SB03} after a period
of time which varies fiber to fiber. The time taken for complete failure
is called the lifetime of the bundle. According to the load sharing
rule,  fiber bundles are being classified into two groups: Equal load-sharing
(ELS) bundles {[}8-17{]} or democratic bundles and local load-sharing
(LLS) bundles \cite{LLS,LLS-th,LLS-rev}. In the ELS models all the
intact fibers equally share the terminal load of a failed fiber, whereas
in LLS model the terminal load gets shared among the intact nearest
neighbors. ELS models show phase transition from partial failure to
total failure at a critical strength ($\sigma_{c}$). The critical
behavior in the failure dynamics of ELS bundles has been solved analytically
\cite{SB01,SBP02} and the universality of the ELS model has been
established \cite{PSB03} recently. However the strength of LLS models
goes to zero \cite{smith-80,pach-93,SB-mod} at the limit of infinite
system size and this does not permit any critical behavior in the
failure process. 

The ELS and LLS models belong to two opposite extremes with respect
to the spatial correlations in stress redistributions. These models
do not incorporate any type of stress gradient among the intact fibers
which is an usual expectation. Therefore a load sharing scheme in
between ELS and LLS should be a realistic approach to study the failure
of heterogeneous materials. Hansen and Hemmer \cite{lamda model}
introduced a `$\lambda$ model' to interpolate between ELS and LLS
models where $\lambda$ is an adjustable stress transfer factor. Although
they conjectured the existence of a critical crossover value $\lambda_{c}$
which separates the mean field (ELS) regime and the short range (LLS)
regime, what would be the exact crossover point was not answered.
A recent approach by Hidalgo et al \cite{variable range} incorporate
both the ELS and LLS mode introducing an effective range of interaction
parameter ($\gamma$) which is actually the power of the stress redistribution
function. They observed crossover behavior in strength variation and
in the avalanche statistics of the failures. Also they determined
the crossover point ($\gamma_{c}$) through the moment analysis of
the cluster size distributions before total failure. 

In this paper we develop a mixed-mode load sharing (MMLS) model which
interpolates the ELS and LLS models correctly. We intend to study
whether this model shows a continuous transition from mean-field (ELS)
behavior to extreme statistics (LLS), or there exists a definite crossover
point. 

We organize this paper as follows: After introduction (section I)
we present our MMLS model in section II. Section III contains the
observations of Crossover behavior through numerical study of the
model. The analysis to determine exact crossover point is given in
section IV. The final section (V) is devoted for discussions including
our conclusions.

\section{The model}

Our mixed-mode load sharing (MMLS) scheme is basically a coupling
of ELS and LLS mode: When a fiber fails, a fraction ($g$) of its
terminal load gets shared among the nearest neighbors of the failed
fiber (LLS rule) and the rest ($1-g$ fraction) is distributed equally
among all the surviving fibers (ELS rule). Here `$g$' is the weight
parameter of the MMLS scheme. Therefore, the model reduces exactly
to ELS model for $g=0$ and for $g=1$, it becomes pure LLS one. As
we have chosen $1-D$ fiber bundle model (with periodic boundary condition),
the number of nearest neighbors is always two. We study the behavior
of the model for the entire range $0\leq g\leq1$ using Monte-Carlo
simulations for step-wise equal load increment \cite{SB01,SBP02,PSB03}
until the total failure of the bundle. During the entire study we
consider uniform (on average) distribution of fiber strength threshold
in the bundle.

\section{The crossover behavior}

\subsection{\noindent Strength of the bundle}

It is known since Daniels \cite{Dan45} that the ELS bundles have
a nonzero strength ($\sigma_{c}$) above which the bundle fails completely.
Recently it has been shown analytically \cite{SB01,SBP02} that for
uniform fiber threshold distribution, bundle's strength approaches
the value $1/4$ as system size goes to infinity. On the other hand
LLS bundles do not have any nonzero strength \cite{smith-80,pach-93,SB-mod}.
In our MMLS model we intend to study the strength variation of bundles
with system size as well as with the weight parameter $g$.

\vskip.1in

\includegraphics[%
  width=2.5in,
  height=2.1in]{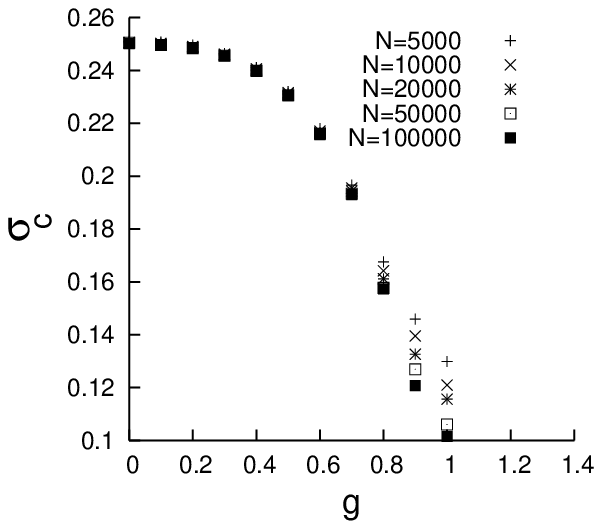}

\textbf{Fig. 1:} {\small The strength of the bundle for different
system sizes ($N$) as a function of the weight parameter $g$.} 

\vskip.1in

As $g$ increases, the bundle becomes weaker due to the short-range
(LLS) interactions. Therefore $\sigma_{c}$ decreases with increasing
$g$ values (Fig. 1). We can see that $\sigma_{c}$ seems to be independent
on system size (dominance of ELS) up to $g=0.7$ and beyond $g=0.8$,
a strong system size dependence (dominance of LLS) appears. This observation
is supported by Fig. 2, where we have shown the logarithmic size dependence
of $\sigma_{c}$. Up to $g=0.7$, the curves eventually become flat
as the system size increases. But for $g\geq0.8$, all the curves
fall (following inclined straight line). Thus the two regimes are
differentiated clearly. 

\vskip .2in

\includegraphics[%
  width=2.5in,
  height=2in]{mixed-log-sigmac.eps}

\textbf{Fig. 2:} {\small The logarithmic size dependence of bundle's
strength for different values of $g$. The straight lines represent
the best fit. }{\small \par}

\subsection{\noindent Avalanche size distribution}

The avalanche size distribution characterizes the fracture process
by reflecting the precursory activities toward complete failure. This
can be related to the acoustic emissions observed in material failure
\cite{AE-97,AE-98,AE_94}. Hemmer and Hansen showed \cite{HH92} analytically
that for ELS models the avalanche size distribution follows an universal
power law with exponent value $-5/2$. But for LLS models the numerically
estimated apparent exponent value is quite larger $4.5$ \cite{HH94}.
Later it has been shown analytically (for flat threshold distributions)
that for LLS model, no universal power-law asymptotics exists \cite{KHH97}. 

\vskip.1in

\includegraphics[%
  width=2.2in,
  height=1.9in]{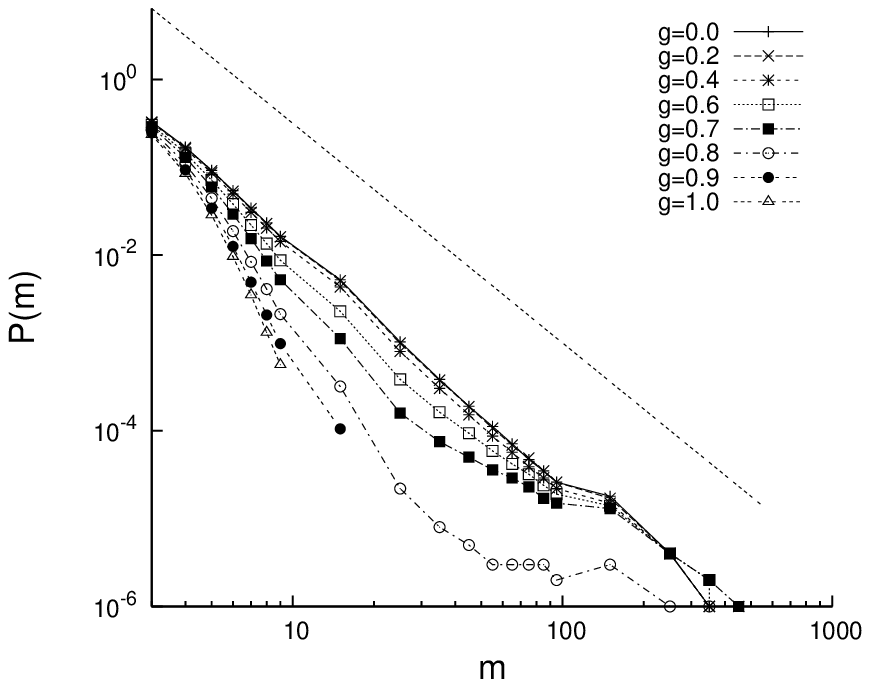}

\textbf{Fig. 3:} A{\small valanche size distribution for different
values of the weight parameter $g$ (averaging over $5000$ configurations
for system size $N=20000$). The dotted line represents mean-field
result having exponent value $-5/2$. Clearly, the upper group of
curves can be fitted by the mean-field power law whereas the lower
group does not show power law at all.}{\small \par}

{\small \vskip.1in}{\small \par}

Here we have measured (Fig. 3) the avalanche size distributions for
different $g$ values. Clearly, two groups of curves appear. The upper
group ($0\leq g\leq0.7$) can be fitted with the mean-field result
($-5/2$) where as the lower group ($0.8\leq g\leq1.0$) show a clear
deviation from the power law.

\subsection{The susceptibility and relaxation time variations}

Recently, the dynamic response parameters, susceptibility ($\chi$)
\cite{RS99,pach-00,SB01,SBP02,Acharyya-96} and relaxation time ($\tau$)
\cite{SB01,SBP02} have been studied in fiber bundle models. The susceptibility
is defined as the number of fibers fail due to an infinitesimal change
of the external stress ($\sigma$) on the bundle and the relaxation
time is the time (number of stress redistributions) the bundle takes
to come to a stable fixed point at an external stress ($\sigma$).
For ELS model, the susceptibility and relaxation time seem to follow
power law with the applied stress and both of them diverge \cite{pach-00,SB01,SBP02,PSB03}
at the critical strength $\sigma_{c}$: $\chi\sim(\sigma_{c}-\sigma)^{-1/2}$
and $\tau\sim(\sigma_{c}-\sigma)^{-1/2}$. However, one can not expect
such scaling behavior in LLS models due to the absence of `critical'
strength. The step-wise equal load increment method \cite{SB01,SBP02}
enables to measure $\chi$ and $\tau$ for different values of $g$
(Fig. 4). The power law behavior (with mean-field exponent $-1/2$)
remains unchanged up to $g=0.7$ and for $g\geq0.8$ the curves do
not follow power laws at all. Thus the susceptibility and relaxation
time variations also suggest a transition from the mean-field to short
range behavior to happen in between $g=0.7$ and $g=0.8$. 

\vskip.2in

\includegraphics[%
  width=1.5in,
  height=1.8in]{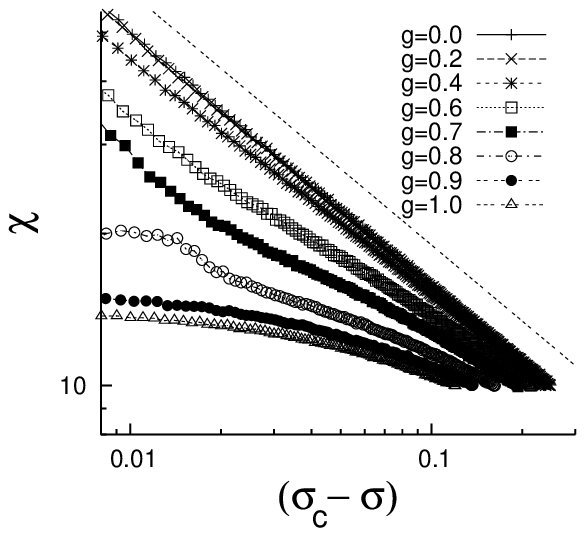}\includegraphics[%
  width=1.5in,
  height=1.8in]{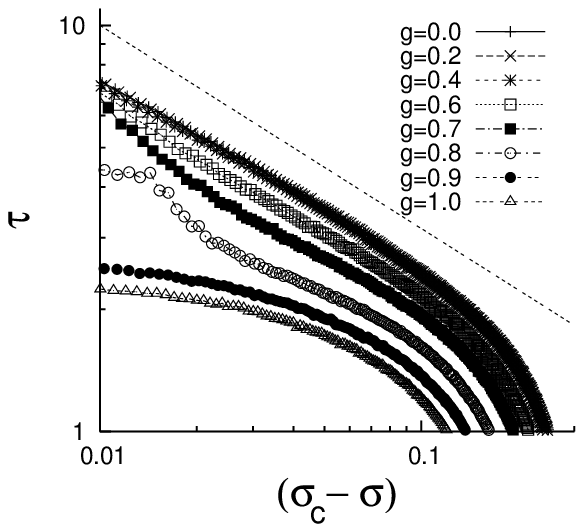}

\textbf{Fig. 4:} {\small The susceptibility ($\chi$) and relaxation
time ($\tau$) variations for different g values.} The bundle contains
$10000$ fibers and the data are averaged over $10000$ configurations.

\subsection{Correlations among the broken fibers}

The breakdown sequence reflects the correlations of the breaking process
\cite{lamda model}. While the ELS model simply ignores the spatial
arrangement of the fibers, LLS model gives much importance on it.
Therefore, as $g$ increases (LLS mode dominates) the breaking process
becomes more and more correlated (Fig. 5). Here also we can identify
two distinct regimes. We cannot see any spatial correlation among
the broken fibers (except near the total failure) up to $g=0.7$,
whereas for $g\geq0.8$ strong correlations (black patch) develop
long before the total failure. 

\vskip.2in

\hskip.5in{\large $g=0.2$}\hskip1.0in {\large $g=0.7$}{\large \par}

\includegraphics[%
  width=1.5in,
  height=1.5in,
  angle=180]{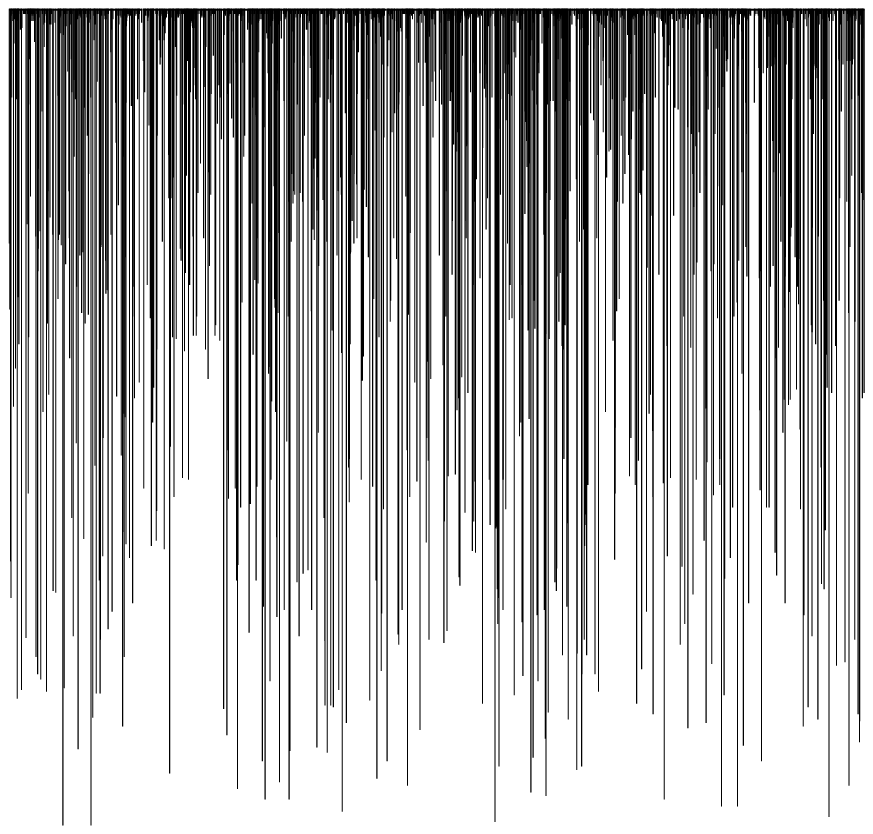}\includegraphics[%
  width=1.5in,
  height=1.5in,
  angle=180]{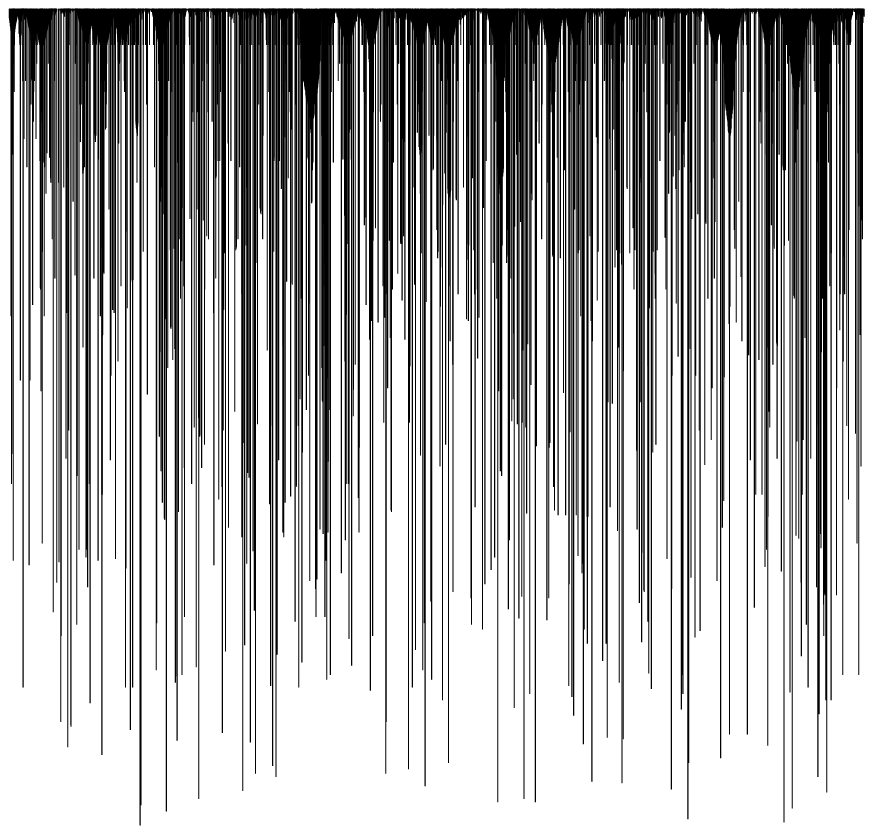}

\vskip.1in

\includegraphics[%
  width=2cm,
  height=2cm]{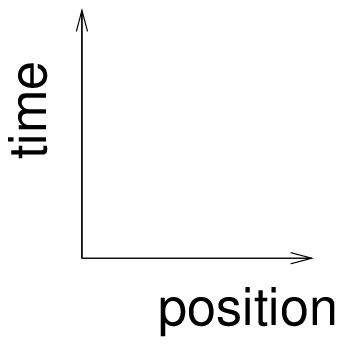}

\vskip.1in

\hskip.5in{\large $g=0.8$}\hskip1.0in {\large $g=1.0$}{\large \par}

\includegraphics[%
  width=1.5in,
  height=1.5in,
  angle=180]{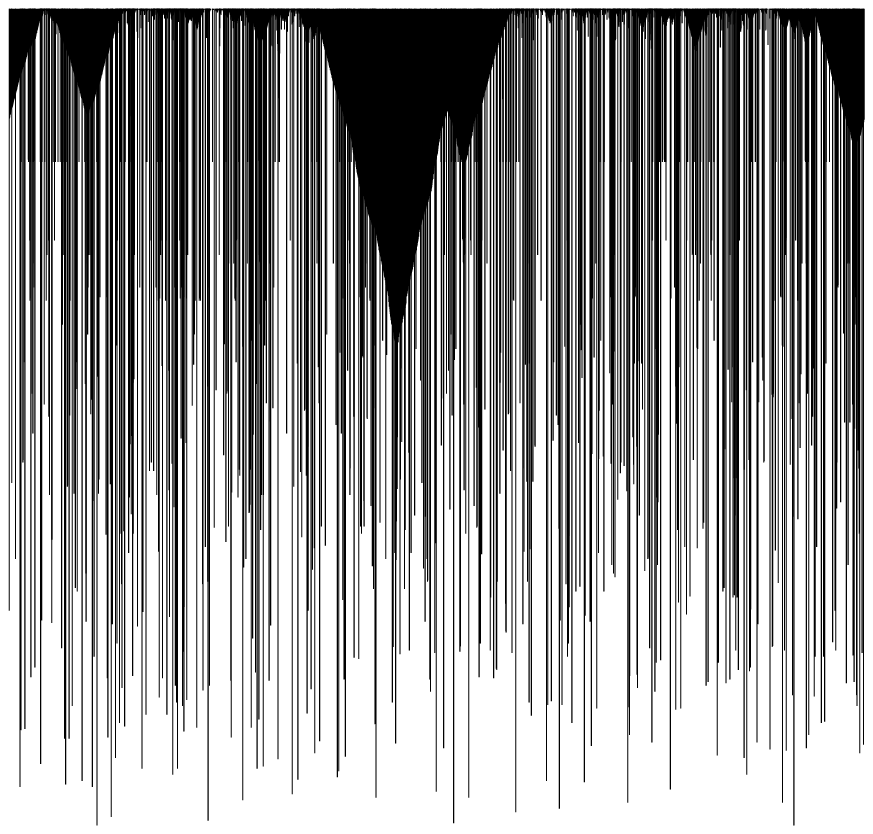}\includegraphics[%
  width=1.5in,
  height=1.5in,
  angle=180]{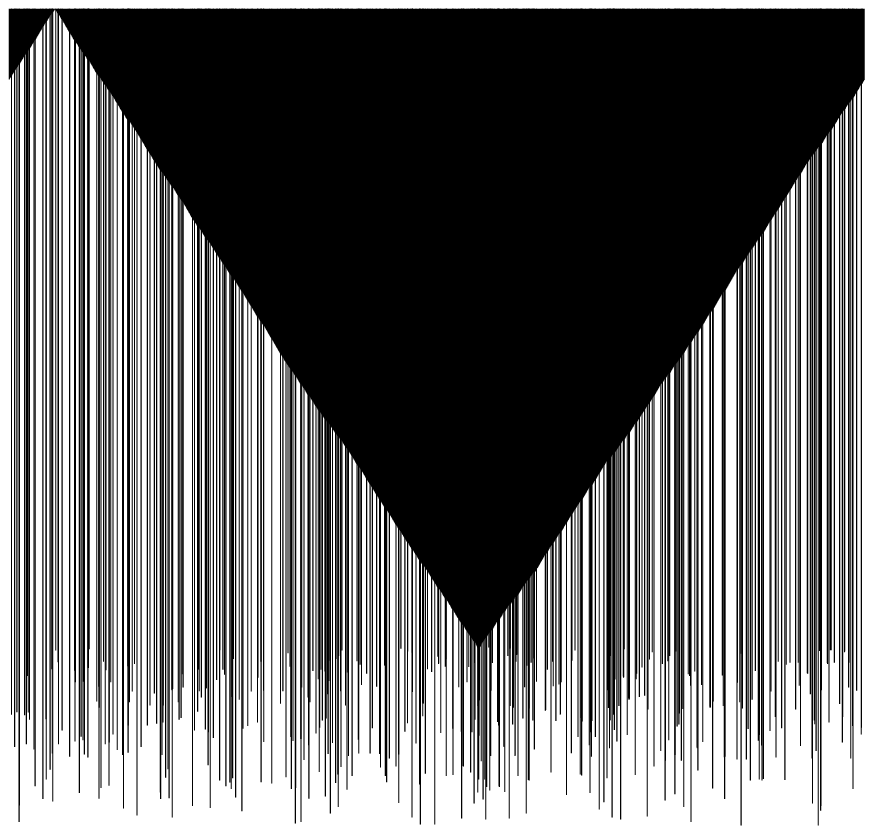}

\textbf{Fig. 5:} {\small The space-time diagram of the breakdown sequence
in MMLS model. The positions of the fibers are marked on the $x$
axis and} $y$ {\small axis is a `time' axis where time indicates
the number of stress redistribution starting from initial loading.
The white color represents intact fibers while the black regions stands
for the broken fibers. }{\small \par}

\section{Determination of the exact Crossover point through cluster moment
analysis }

The fracture process can also be characterized by analyzing the clusters
of broken fibers just before complete failure \cite{Phase T,variable range}.
The size distributions of the clusters ($n(s)$ vs. $s$) are shown
(Fig. 6) for different values of $g$. Although the distributions
appear as two groups, it is not possible to identify the exact crossover
point from this. Therefore we go for the moment analysis: the $k$-th
moment of the cluster distributions is defined \cite{variable range}
as\begin{equation}
m_{k}=\int s^{k}n(s)ds\label{eq:1}\end{equation}

Clearly the zero-th moment ($m_{0}$) gives the total number clusters
and the first moment gives the total number of broken fibers. We can
get the average cluster size dividing the second moment ($m_{2}$)
by first moment ($m_{1}$). 

\vskip.2in

\includegraphics[%
  width=3in,
  height=2.5in]{mixed-ns-new.eps}

\textbf{Fig. 6:} {\small Cluster size distributions of broken fibers
(just before complete failure) for different g values (averaging over
$5000$ samples for $N=20000$). }{\small \par}

\vskip.2in

\includegraphics[%
  width=1.5in,
  height=1.8in]{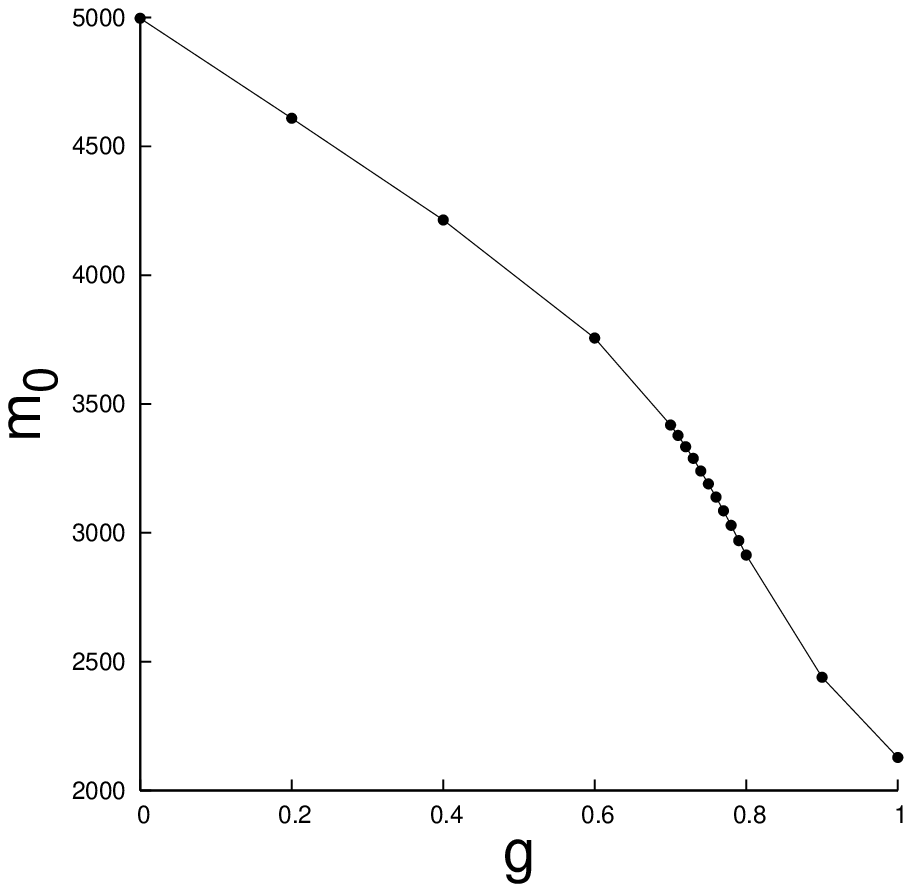}\includegraphics[%
  width=1.5in,
  height=1.8in]{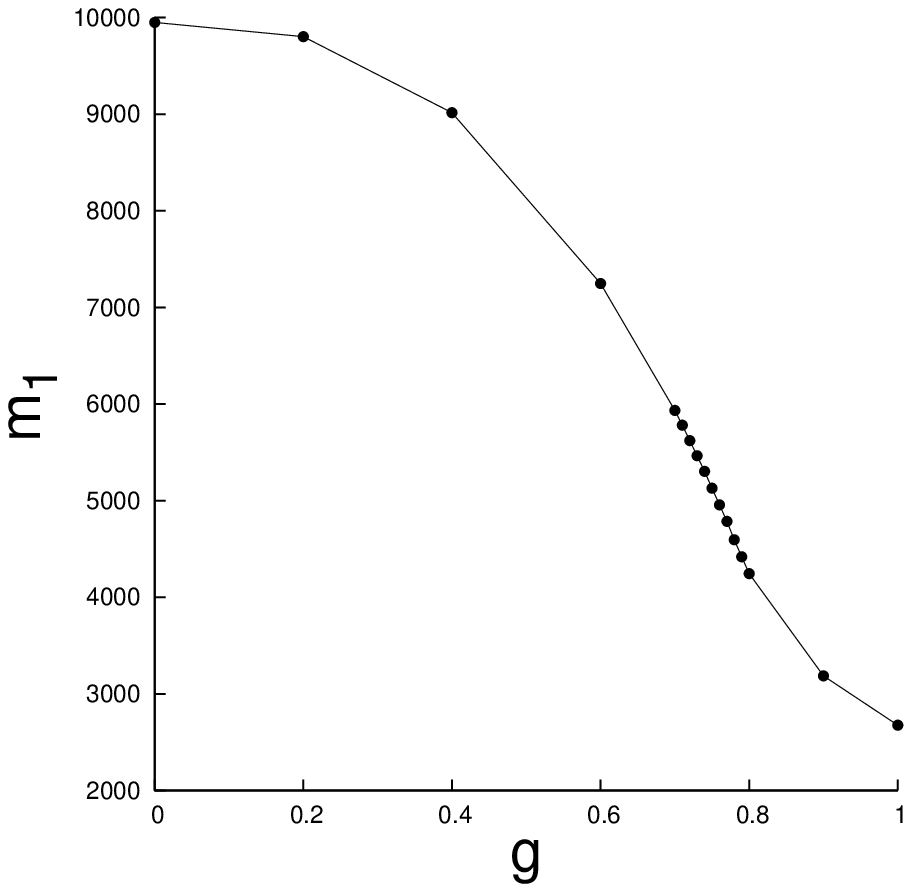}

\includegraphics[%
  width=1.5in,
  height=1.8in]{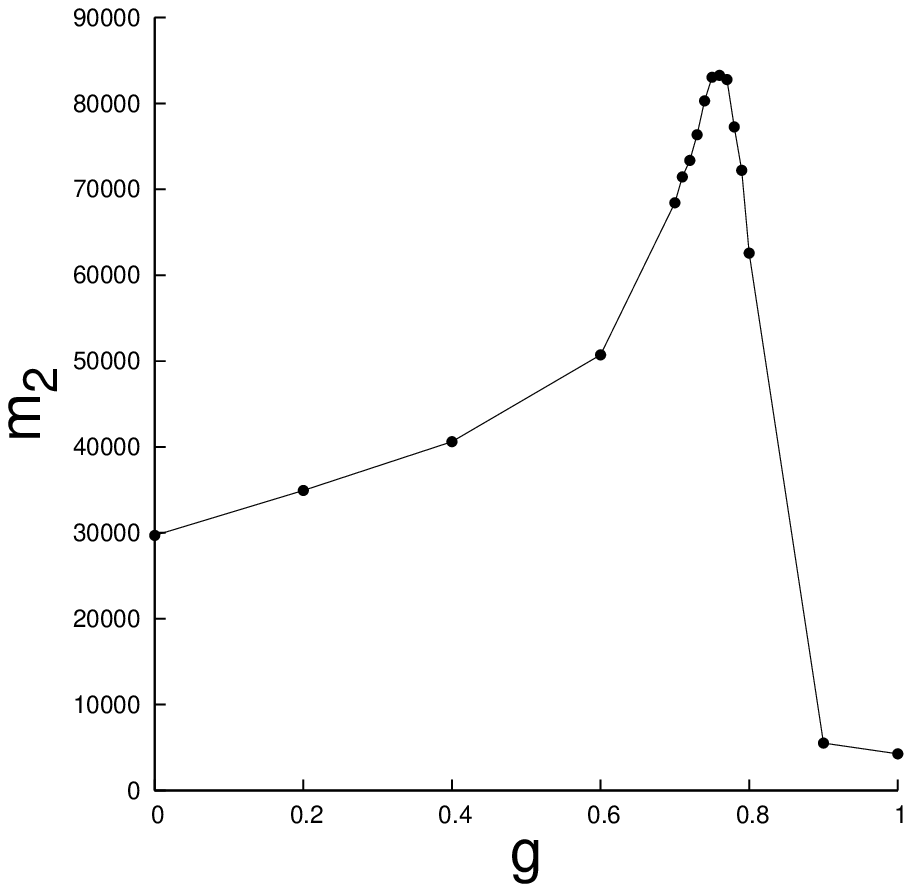}\includegraphics[%
  width=1.5in,
  height=1.8in]{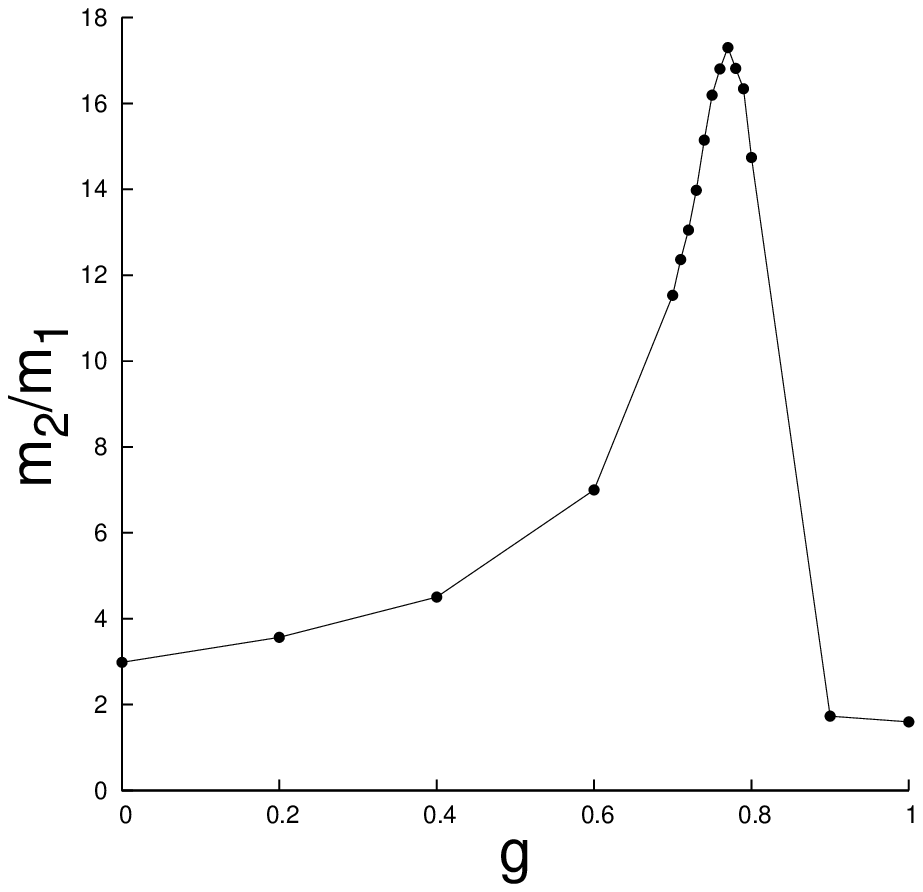}

\textbf{Fig. 7:} {\small The moments of the cluster size distributions
as a function of the weight parameter g (averaging over $5000$ samples
for $N=20000$) .} 

\vskip.2in

In case of pure ELS mode ($g=0$), we have only long-range interaction
and the clusters are randomly distributed within the lattice. As $g$
increases the stress redistribution becomes more and more localized
in the neighborhood of the failed fibers and a few isolated crack
can trigger the complete rupture through growth and coalescence mechanism.
Therefore the pure ELS mode can store the maximum crack (cluster)
and this capacity should decrease with the increase of $g$. We can
see (Fig. 7) that both $m_{0}$ and $m_{1}$ decrease with increasing
$g$ value and they fall drastically in between $g=0.7$ and $g=0.8$.
This crossover is much robust in case of $m_{2}$ and the average
cluster size ($m_{2}/m_{1}$), both of which show a sharp peak, which
indicates the dominance of LLS mode over the ELS mode \cite{variable range}. 

\vskip.2in

\includegraphics[%
  width=2.5in,
  height=2.2in]{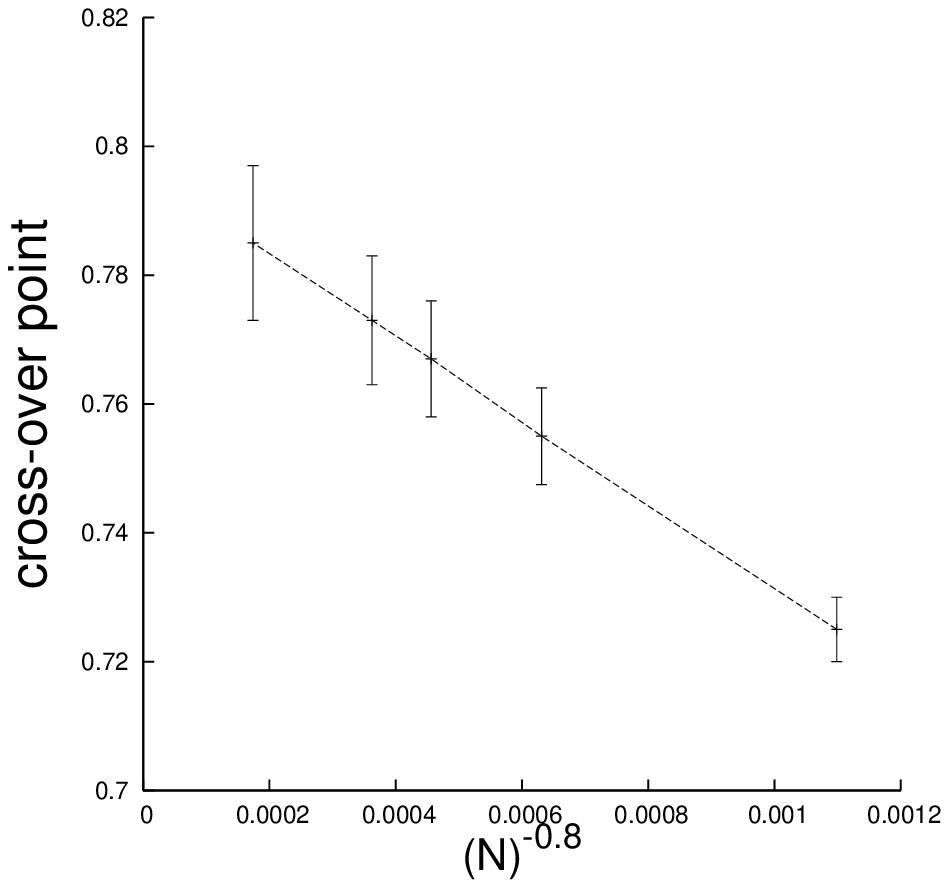}

\textbf{Fig. 8:} {\small System size dependence of the crossover point.} 

\vskip.1in

To check how the crossover point changes its position with system
size, we have done the similar cluster moment analysis for several
system sizes. We observe a weak system size dependence of the peak
position i.e., the crossover point (Fig. 8). With proper extrapolation
we determine the crossover point ($g_{c}$) to be at $g=0.79\pm0.01$
for the infinite system size.

\section{Conclusion}

The fracture and breakdown of loaded materials is basically a cooperative
phenomenon guided by the load redistribution mechanism. Here `crack'
opens up when an element (fiber) fails after external loading. This
single fiber failure should affect the neighbors much than the distant
elements (like in electric Fuse Models \cite{books}). Therefore a
high stress concentration (after the load redistribution) around a
crack (failed fiber) is a natural expectation. The ELS models do not
incorporate any spatial correlations and exhibit perfect democracy
(mean-field), whereas the LLS models confine themselves within the
nearest neighbor interactions. In this situation attempts \cite{lamda model,variable range}
to study the failure behavior in between ELS and LLS regimes, would
be most welcome. Also, a recent experiment on loaded wood-fiber \cite{wood-expt}
demands an intermediate load sharing scheme to explain the observed
strength variation. The `$\lambda$ model' \cite{lamda model} becomes
a LLS model at $\lambda=1$. But it cannot be reduced to a pure ELS
model at $\lambda=0$, as the neighbors of the just broken fibers
become `immunized' against failure. Although the `variable range of
interaction' model \cite{variable range} determines the exact crossover
point, it remains silent about the system size dependence of crossover
point, which is nevertheless an important issue. 

Our mixed-mode load sharing (MMLS) model exactly reduces to ELS model
at $g=0$ and to LLS model at $g=1$. We establish numerically that
the MMLS model in one dimension shows a distinct crossover behavior
from mean-field to short-range interaction. The strength ($\sigma_{c}$)
variation of the bundle with system size, the avalanche statistics
and the failure dynamics (susceptibility and relaxation time) suggest
that the crossover point ($g_{c}$) must be in between $g=0.7$ and
$g=0.8$. The cluster size analysis determines the exact crossover
point in one dimension for several system sizes and a proper extrapolation
suggests the crossover point to be $g_{c}=0.79\pm0.01$ at the limit
of infinite system size. For $g<g_{c}$ the model exhibits critical
behavior (supported by the power laws) for the dominance of ELS mode.
But the fluctuations suppress any critical behavior after $g=g_{c}$,
where extreme statistics \cite{books} dominates. We should mention
that as the ultimate strength ($\sigma_{c}$) of the bundle continuously
decreases with the increasing $g$ value, we cannot exclude the possibility
of different critical behavior for $g=0$, $0<g\leq g_{c}$ and $g>g_{c}$
in higher dimensions, like in case of $2-D$ Ising systems with disorder
\cite{Stinch}. Therefore we expect this crossover behavior in MMLS
model to be more prominent in higher dimensions. 

\vskip.2in

\textbf{Acknowledgment}: We are grateful to Dr. M. Kloster for useful
comments. S. P. thanks the Norwegian Research Council, NFR for the
funding through a strategic university program.


\begin{thebibliography}{10}
\bibitem{books}H. J. Herrmann and S. Roux (Eds), \emph{Statistical Models for the
Fracture of Disordered Media}, North Holland, Amsterdam (1990); B.
K. Chakrabarti and L. G. Benguigui, \textit{Statistical Physics of
Fracture and Breakdown in Disorder Systems}, Oxford Univ. Press, Oxford
(1997); M. Sahimi, \emph{Heterogeneous Materials II: Nonlinear and
Breakdown Properties}, Springer-Verlag, New York (2003).
\bibitem{Peirce}F. T. Peirce, J. Textile Inst. \textbf{17}, T355-368 (1926). 
\bibitem{Dan45}H. E. Daniels, Proc. R. Soc. London A \textbf{183} 405 (1945). 
\bibitem{BD-58}B. D. Coleman, J. Appl. Phys. \textbf{27}, 862 (1956); B. D. Coleman,
Trans. Soc. Rheol. \textbf{1}, 153 (1957); B. D. Coleman, Trans. Soc.
Rheol. \textbf{2}, 195 (1958).
\bibitem{Phoenix00}W. I. Newman and S. L. Phoenix, Phys. Rev. E \textbf{63}, 021507 (2000). 
\bibitem{Roux-00}S. Roux, Phys. Rev. E \textbf{62}, 6164 (2000); R. Scorretti, S. Ciliberto
and A. Guarino, Europhys. Lett. \textbf{55}, 626 (2001). 
\bibitem{SB03}S. Pradhan and B. K. Chakrabarti, Phys. Rev. E \textbf{6}7, 046124
(2003).
\bibitem{HH92}P. C. Hemmer and A. Hansen, J. Appl. Mech. \textbf{59} 909 (1992).
\bibitem{HH94}A. Hansen and P. C. Hemmer, Phys. Lett. A \textbf{184} 394 (1994).
\bibitem{KHH97}M. Kloster, A. Hansen and P. C. Hemmer, Phys. Rev. E \textbf{56} 2615
(1997). 
\bibitem{D Sornet}D. Sornette, J. Phys. A \textbf{22} L243 (1989); D. Sornette, J. Phys.
I (France) \textbf{2} 2089 (1992); A. T. Bernardes and J. G. Moreira,
Phys. Rev. B \textbf{49} 15035 (1994).
\bibitem{Phase T}S. Zapperi, P. Ray, H. E. Stanley and A. Vespignani, Phys. Rev. Lett.
\textbf{78} 1408 (1997); 
\bibitem{pach-00}Y. Moreno, J. B. Gomez and A. F. Pacheco, Phys. Rev. Lett. \textbf{85}
2865 (2000).
\bibitem{RS99}R. da Silveira, Am. J. Phys. \textbf{67} 1177 (1999). 
\bibitem{SB01}S. Pradhan and B. K. Chakrabarti, Phys. Rev. E \textbf{65}, 016113
(2001); 
\bibitem{SBP02}S. Pradhan, P. Bhattacharyya and B. K. Chakrabarti, Phys. Rev. E \textbf{66},
016116 (2002); 
\bibitem{PSB03}P. Bhattacharyya, S. Pradhan and B. K. Chakrabarti, Phys. Rev. E \textbf{6}7,
046122 (2003);
\bibitem{LLS-th}D. G. Harlow and S. L. Phoenix, J. Composite Mater. \textbf{12}, 314
(1978); R. L. Smith, Adv, Appl. Prob. \textbf{15}, 304 (1982). 
\bibitem{LLS}S. L. Phoenix, Adv. Appl. Prob. \textbf{11}, 153 (1979); R. L. Smith
and S. L. Phoenix, J. Appl. Mech. \textbf{48}, 75 (1981).
\bibitem{LLS-rev}S. L. Phoenix and R. L. Smith, J. Solid Struct. \textbf{19}, 479 (1983). 
\bibitem{smith-80}R. L. Smith, Proc. R. Soc. London \textbf{A 372}, 539 (1980). 
\bibitem{pach-93}J. B. Gomez, D. Iniguez and A. F. Pacheco, Phys. Rev. Lett. 71, 380
(1993).
\bibitem{SB-mod}S. Pradhan and B. K. Chakrabarti, Int. J. Mod. Phys. B \textbf{17}
5565 (2003). 
\bibitem{lamda model}A. Hansen and P. C. Hemmer, Trends in Stat. Phys. \textbf{1}, 213
(1994). 
\bibitem{variable range}R. C. Hidalgo, Y. Moreno, F. Kun and H. J. Herrmann, Phys. Rev. E
\textbf{65} 046148 (2002).
\bibitem{AE-97}A. Garcimartin, A. Guarino, L. Bellon and S. Ciliberto, Phys. Rev.
Lett. \textbf{79} 3202 (1997). 
\bibitem{AE-98}A. Guarino, A. Garcimartin and S. Ciliberto, Eur. Phys. J. B \textbf{6}
13 (1998). 
\bibitem{AE_94}A. Petri, G. Paparo, A. Vespignani, A. Alippi and M. Costantini, Phys.
Rev. Lett \textbf{73} 3423 (1994). 
\bibitem{Acharyya-96}M. Acharyya and B. K. Chakrabarti, Physica A \textbf{224} 254 (1996). 
\bibitem{wood-expt}G. D. Langer, R. C. Hidalgo, F. Kun, Y. moreno, S. Aicher and H. J.
Herrmann, Physica A \textbf{325} 547 (2003).
\bibitem{Stinch}R. Stnchcombe, in C. Domb and J. L. Lebowitz, \emph{Phase Transition
and Critical Phenomena}, Vol. \textbf{7} 151 (1983).\end{thebibliography}
\end{document}